\documentclass[aps,prb,twocolumn,superscriptaddress,showpacs]{revtex4}
\usepackage{microtype}
\usepackage[dvips]{graphicx,color}
\usepackage{amsmath}
\begin{document}
%%%%%%%%%%%%%%%%%%%%%%%%%%%%%%%%%%%
\title{Universal van der Waals-type interactions in rattler containing cage materials}
%%%%%%%%%%%%%%%%%%%%%%%%%%%%%%%%%%%
%%%Authhors
\author{Jiazhen Wu}
\thanks{wujzphystu@gmail.com}
\affiliation{WPI-AIMR, Tohoku University, Sendai, 980-8577, Japan\\
}
%%---------------------------
\author{Jingtao Xu}
\affiliation{Ningbo Institute of Materials Technology $\&$ Engineering, Chinese Academy of Sciences, Ningbo, China\\
}
%%---------------------------
\author{Katsumi Tanigaki}
\thanks{tanikagi@sspns.phys.tohoku.ac.jp}
\affiliation{WPI-AIMR, Tohoku University, Sendai, 980-8577, Japan\\
}
\affiliation{Graduate School of Science, Tohoku University, Sendai 980-8578, Japan\\
}
%%%****************************************

%%=== Abstract
\begin{abstract}
Rattling motion of fillers in cage materials has been of great interest for their import roles in superconductivity and thermoelectric applications.
The standing waves of the rattling oscillations are normally lower in energy than the propagating waves of the acoustic phonons, thus exert large influences on the configuration of phonon dispersions as well as the associated thermal and electrical properties.
Although it has been extensively studied, the origin of the low energy soft modes is still not clear.
In the present paper, we show that van der Waals-type interactions are predominant between fillers and their surrounding cage frameworks, which explains the origin of the low energy modes in cage materials as a universal rule.
Mass, free space and chemical environment of guest atoms are shown to be the most important factors to determine the three dimensional van der Waals-type interactions.
%%Our discussion is based on the the characteristic energies of rattlers and the associated space inside cages as well as a harmonic vibration approximation.
The present work is mainly focused on type-I clathrates, skutterudites and pyrochlores.
%%our discussion is made with a harmonic lattice approximation on the basis of a summary the characteristic energies of different rattlers with different cages and the associated space factor with a harmonic approximation.
\end{abstract}
%%%=======================
%%\pacs{***}

\maketitle
%%%********************************************************************
\section*{Introduction}%%The introduction is a little bit hard for reader to understand, no details? (noted on 0810)
\begin{figure}[htbp]
\centering
\includegraphics*[width=0.85\linewidth]{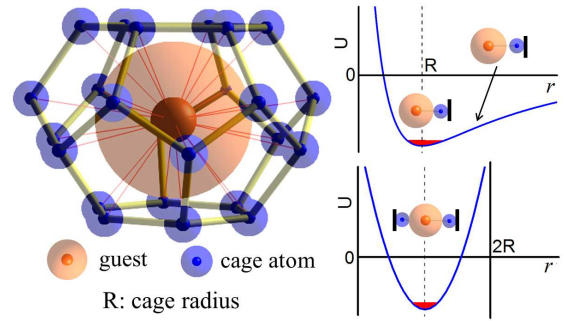}
\caption
{
Left: a sketched rattler containing cage.
The cage atoms (hosts) are bonded via covalent interaction leaving enough room for the guest to move freely.
The atomic radius of the guest atom is enlarged to illustrate the relatively large interspace.
Right: sketched van der Waals potentials for a pair of guest-host atoms and a set of host-guest-host atoms.
The later case is a simplified picture for a rattler containing cage.
}
\label{Fig1}
\end{figure}

%%--Rattling in cage materials, ALE modes, and its importance
The rattling phenomenon might be first described by A. J. Sievers and S. Takeno in 1965, when they observed an anomalous peak of Li in KBr:LiBr at low frequencies.~\cite{SieversPR1965}
Recently, rattling behavior has been commonly observed in filler containing cage materials, such as Al$_{10}$V-type intermetallides,~\cite{CaplinPRL1973} brownmillerite,~\cite{RykovPBCM2004} skutterudite,~\cite{SalesPRB1997} pyrochlore~\cite{HiroiPRB2007} and clathrate~\cite{SalesPRB2001,IversenNaturematerials2008}.
Different from the Li-rattlers, the fillers (or guest atoms) in cage compounds are accommodated inside a periodic array of cages and can be quantitatively described.
Fig.~\ref{Fig1} shows a rattler containing cage, where the cage atoms are connected by covalent bonds, providing a strong wall and an oversized room for the guest atom.
Since a guest atom is only weakly interacted with the surrounding cage atoms, it normally shows an anomalous low energy (ALE) frequency and a large atomic displacement parameter (ADP).
Although showing anharmonicity, the ALE modes can be approximately described by the Einstein model, where guest atoms are treated as Einstein oscillators with a characteristic energy of $\omega_{E}$.
Rattling ALE phonon modes are scientifically important for: (1) their coupling with conduction electrons, giving rise to superconductivity~\cite{HiroiPRB2007,YamanakaIC2000} and an modification of electron effective mass~\cite{XuPRB2010,WuPRB2014}; and (2) their coupling with propagating phonons, leading to an enhancement of scattering probability and consequent low thermal conductivity for thermoelectric applications~\cite{CohnPRL1999,TadanoPRL2015}.

%%--Experimental techniques and ALE modes----->origin is not clear----
In the passed decades, the rattling phonon modes have been both experimentally and theoretically studied.
The direct evidence has been provided by the phonon dispersion relationship of a type-I clathrate Ba$_{8}$Ga$_{16}$Ge$_{30}$, derived from inelastic neutron scattering measurements.~\cite{IversenNaturematerials2008}
The rattling phonon modes are optical-branch-like at around 5\,meV, lower than the top of acoustic phonon branches, giving rise to ALE peaks in the derived phonon density of states (PDOS).
The ALE vibration modes have also been detected and discussed by Raman spectroscopy,~\cite{KumePRL2003} optical conductivity,\cite{MoriPRB2009} heat capacity (HC),~\cite{Takabatake2014} and temperature dependent ADP obtained from crystallographic refinement of x-ray/neutron diffraction data~\cite{SalesPRB1997}.
%%not just cite paper in clathrate fields
These experiment results have established the validity of the Einstein model for the description of the ALE modes; however exceptions were found for guest off-centered modes~\cite{SalesPRB2001,SuekuniPRB2008} which were rather anharmonic and beyond the scope of the Einstein model and the present discussion.
In addition, many theoretical investigations have been performed as well, aiming at unveiling the rattling phonon associated phonon static and dynamic properties.~\cite{TadanoPRL2015,Nakayama}
Nevertheless the origin of the low energy modes was not clear.
%%phonon scattering mechanism

%%%-----------------------
%%--Our work
In our previous work,~\cite{WuPRB2016} we addressed a van der Waals-type interaction between a filler and its surrounding cage framework as the origin of the rattling ALE modes by focusing only on the 6d parallel modes in type-I clathrates.
In the present study we extend our scope to the other two modes in type-I clathrates (2a and 6d perpendicular modes, refer to Fig.\,\ref{Fig2} for description of the modes) and the rattling modes in skutterudites and pyrochlores to unveil that the three dimensional (3D) van der Waals interaction is universal for rattler containing cage materials.
Force constants, derived from the characteristic energies of guest atoms, are shown to vary exponentially with the free space of guest atoms ($R_{\mathrm{free}}$) for each type of rattling mode.
Different exponential parameters are obtained for different types of rattling modes, indicating that, besides space parameters discussed previously,~\cite{WuPRB2016} chemical environment around guest atoms is another important factor that influences the guest-host interactions.
The 3D van der Waals-type interaction, which coexists with the strong interatomic covalent interaction of framework atoms, significantly modifies the physical properties of a cage material via phonon-phonon and electron-phonon interactions, and provides a platform to study superconductivity and thermoelectricity.
%%local vibration associated physical properties.
%%%===****************************************************

\section*{Analyses method}
%%%======================
\subsection{Vibration modes and data collection}
\begin{figure*}[htbp]
\centering
\includegraphics*[width=0.9\linewidth]{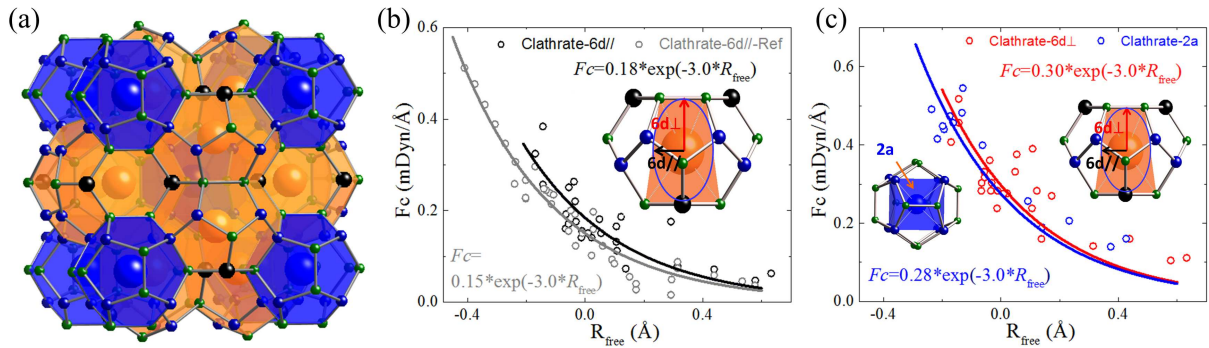}
\caption
{
(a)\,lattice structure of type-I clathrate.
The cage framework is mainly composed of Si/Ge/Sn and the colors of the atoms indicate different crystallographic sites.
The guest atoms are alkali metals or alkaline earth metals.
There are two types of cages, dodecahedral and tetrakaidecahedral cages, which are shown by blue and orange colors, respectively.
(b,c)\,The relationship between Fc and $R_{\mathrm{free}}$ for different vibration modes in type-I clathrates.
The closest interspaces between guest and cage atoms are shown in the inset of the figures for both types of cages.
The arrows show anisotropic vibration modes (6d modes) in the tetrakaidecahedral cage.
The lines are fitting results by employing exponential functions.
The original data are derived from Refs.,\,\cite{ChristensenCM2007,Johnsen2,Johnsen,Christensen4,ChristensenJACS2006,FalmbiglIC2013,ZeiringerActaMaterialia2011,BentienPRB2005,MelnychenkoPRB2007,ChristensenDT2010,NasirJPCM2009,Bentien,FalmbiglInter2013,Falmbigl,SChristensen,Tanaka,Stefanoski,RoudebushIC2012,KaltzoglouJMC2008,Shimizu}
and the data with grey color is taken from our previous work.\,\cite{WuPRB2016}
}
\label{Fig2}
\end{figure*}

%%Vibration modes
The current research is concerning the ALE vibration modes of the rattler in cage compounds.
The situation for skutterudite and pyrochlore is simpler, because they are composed of single type of cages and have only one guest-vibration mode (as shown in Fig.\,\ref{Fig3}).
However, type-I clathrate, which contains two type of cages, is much more complicated.
As shown in Fig.\,\ref{Fig2}, there are three guest modes: 2a, 6d$\parallel$ and 6$\perp$ modes, which are shown by the blue cage, black arrow and red arrow in the orange cage, respectively.
Here 2a and 6d represent the crystallographic sites on the cage centers, $\parallel$ and $\perp$ are the directions with respect to the six-atom ring of the tetrakaidecahedral cage.
%%%======================

%%Data source: Cp, Raman, INS, ADP; from Refs.     ADP good for higher energy modes.
Different ALE modes have different characteristic energies ($\omega_{E}$), depending on the chemical environment and the relative free space that rattlers can move.
In principle, the vibration energies can be directly detected by INS, Raman spectroscopy and heat capacity measurements.
However, if $\omega_{E}$ is not low enough, and close to the top of an acoustic phonon branch, where the acoustic phonon becomes less propagative (the acoustic phonon branch becomes flat at the Brillouin zone edge), $\omega_{E}$ cannot be precisely obtained from these experiment techniques, because the contributions from van Hove singularities cannot be extracted.
We have exemplified the situation in the heat capacity analyses for K$_{8}$Ga$_{8}$Sn$_{38}$, where the number of ALE modes obtained from data fitting exceeds what we expect from the number of guest atoms, showing clearly the contributions from van Hove singularities.~\cite{WuPRB2016}
In order to obtain a correct value for $\omega_{E}$, temperature dependent ADP of a guest atom is useful.~\cite{SalesPRB1997}
%%Temperature dependent ADP is another powerful tool to measure $\omega_{E}$~\cite{ADPScience}, and importantly, the influence of van Hove singularities is not included.
Therefore, the $\omega_{E}$'s in the present analyses are mainly derived from ADP data from literatures (Refs.~\cite{ChristensenCM2007,Johnsen2,Johnsen,Christensen4,ChristensenJACS2006,FalmbiglIC2013,ZeiringerActaMaterialia2011,BentienPRB2005,MelnychenkoPRB2007,ChristensenDT2010,NasirJPCM2009,Bentien,FalmbiglInter2013,Falmbigl,SChristensen,Tanaka,Stefanoski,RoudebushIC2012,KaltzoglouJMC2008,Shimizu,SchnellePRB2008,XuJMCA2015,XuJEM2012}) for the three modes of type-I clathrates and the guest modes of skutterudites.
The $\omega_{E}$'s for pyrochlores are taken from heat capacity data reported by Hiroi \textit{et. al.}~\cite{HiroiJPSJ2012}
The data are summarized in table\,\ref{SourceData}.

\begin{table*}[htbp]
\centering
\caption{The values of $\omega_{E}$ for the present analyses.}
%%%---
\begin{center}
  \begin{tabular}{cccccccccc}
  \hline \hline
compounds&vibration&$\omega_{E}$&Ref. &\quad& compounds&vibration&$\omega_{E}$&Ref. \\
         &modes    &(K)         &     &\quad&          &modes    &(K)         &      \\ \hline
 \textbf{Type-I Clathrate} & & & &\quad& & & & \\ \hline
Ba$_{8}$Al$_{16}$Si$_{30}$&Ba(6d$\parallel$)&68 &ADP\cite{ChristensenDT2010}&\quad&Ba$_{8}$Al$_{16}$Ge$_{30}$&Ba(6d$\parallel$)&69 &ADP\cite{ChristensenCM2007}\\ \hline
Ba$_{8}$Al$_{16}$Si$_{30}$&Ba(6d$\perp$)    &92 &ADP\cite{ChristensenDT2010}&\quad&Ba$_{8}$Al$_{16}$Ge$_{30}$&Ba(6d$\perp$)    &85 &ADP\cite{ChristensenCM2007}\\ \hline
Ba$_{8}$Al$_{16}$Si$_{30}$&Ba(2a)           &111&ADP\cite{ChristensenDT2010}&\quad&Ba$_{8}$Al$_{16}$Ge$_{30}$&Ba(2a)           &101&ADP\cite{ChristensenCM2007}\\ \hline
Ba$_{8}$Zn$_{7}$Si$_{39}$ &Ba(6d$\parallel$)&80 &ADP\cite{NasirJPCM2009}    &\quad&Ba$_{8}$Ni$_{6}$Ge$_{40}$ &Ba(6d$\parallel$)&75 &ADP\cite{ChristensenDT2010}\\ \hline
Ba$_{8}$Zn$_{7}$Si$_{39}$ &Ba(6d$\perp$)    &90 &ADP\cite{NasirJPCM2009}    &\quad&Ba$_{8}$Ni$_{6}$Ge$_{40}$ &Ba(6d$\perp$)    &85 &ADP\cite{ChristensenDT2010}\\ \hline
Ba$_{8}$Ga$_{16}$Si$_{30}$&Ba(6d$\parallel$)&66 &ADP\cite{ChristensenDT2010}&\quad&Ba$_{8}$Ni$_{6}$Ge$_{40}$ &Ba(2a)           &103&ADP\cite{ChristensenDT2010}\\ \hline
Ba$_{8}$Ga$_{16}$Si$_{30}$&Ba(6d$\perp$)    &92 &ADP\cite{ChristensenDT2010}&\quad&Ba$_{8}$Cu$_{6}$Ge$_{40}$  &Ba(6d$\parallel$)&63 &ADP\cite{Johnsen}\\ \hline
Ba$_{8}$Ga$_{16}$Si$_{30}$&Ba(2a)           &112&ADP\cite{ChristensenDT2010}&\quad&Ba$_{8}$Cu$_{6}$Ge$_{40}$  &Ba(6d$\perp$)    &85 &ADP\cite{Johnsen}\\ \hline
Ba$_{8}$Rh$_{2.4}$Si$_{43.6}$&Ba(6d$\parallel$)&99&ADP\cite{FalmbiglInter2013}&\quad&Ba$_{8}$Cu$_{6}$Ge$_{40}$&Ba(2a)           &106&ADP\cite{Johnsen}\\ \hline
Ba$_{8}$Rh$_{2.4}$Si$_{43.6}$&Ba(6d$\perp$)    &115 &ADP\cite{FalmbiglInter2013}&\quad&Ba$_{8}$Zn$_{8}$Ge$_{38}$&Ba(6d$\parallel$)&62 &ADP\cite{Christensen4}\\ \hline
Ba$_{8}$Ag$_{5}$Si$_{41}$&Ba(6d$\parallel$)&81 &ADP\cite{ZeiringerJJAP2011}&\quad&Ba$_{8}$Zn$_{8}$Ge$_{38}$     &Ba(6d$\perp$)    &98 &ADP\cite{Christensen4}\\ \hline
Ba$_{8}$Ag$_{5}$Si$_{41}$&Ba(6d$\perp$)    &108&ADP\cite{ZeiringerJJAP2011}&\quad&Ba$_{8}$Zn$_{8}$Ge$_{38}$     &Ba(2a)           &118&ADP\cite{Christensen4}\\ \hline
Ba$_{8}$Ni$_{3}$Si$_{43}$&Ba(6d$\parallel$)&91 &ADP\cite{Falmbigl}&\quad&Ba$_{8}$Ga$_{16}$Ge$_{30}$&Ba(6d$\parallel$)&60 &ADP\cite{ChristensenJACS2006}\\ \hline
Ba$_{8}$Ni$_{3}$Si$_{43}$&Ba(6d$\perp$)    &102&ADP\cite{Falmbigl}&\quad&Ba$_{8}$Ga$_{16}$Ge$_{30}$&Ba(6d$\perp$)    &84 &ADP\cite{ChristensenJACS2006}\\ \hline
Na$_{8}$Si$_{46}$&Na(6d$\parallel$)&110 &ADP\cite{Stefanoski}&\quad&     Ba$_{8}$Ga$_{16}$Ge$_{30}$&Ba(2a)           &108&ADP\cite{ChristensenJACS2006}\\ \hline
Na$_{8}$Si$_{46}$&Na(6d$\perp$)    &147 &ADP\cite{Stefanoski}&\quad&Ba$_{8}$Rh$_{1.2}$Ge$_{42.8}$&Ba(6d$\parallel$)&70 &ADP\cite{FalmbiglIC2013}\\ \hline
Na$_{8}$Si$_{46}$&Na(2a)           &170 &ADP\cite{Stefanoski}&\quad&Ba$_{8}$Rh$_{1.2}$Ge$_{42.8}$&Ba(6d$\perp$)    &88 &ADP\cite{FalmbiglIC2013}\\ \hline
Sr$_{8}$Al$_{11}$Si$_{35}$&Sr(6d$\parallel$)&67 &ADP\cite{Takabatake2014}&\quad&Ba$_{8}$Ag$_{6}$Ge$_{40}$  &Ba(6d$\parallel$)&60 &ADP\cite{Johnsen}\\ \hline
Sr$_{8}$Al$_{11}$Si$_{35}$&Sr(6d$\perp$)    &125&ADP\cite{Takabatake2014}&\quad&Ba$_{8}$Ag$_{6}$Ge$_{40}$  &Ba(6d$\perp$)    &78 &ADP\cite{Johnsen}\\ \hline
Sr$_{8}$Al$_{6}$Ga$_{10}$Si$_{30}$&Sr(6d$\parallel$)&54 &ADP\cite{Shimizu}&\quad&Ba$_{8}$Ag$_{6}$Ge$_{40}$ &Ba(2a)           &107&ADP\cite{Johnsen}\\ \hline
Sr$_{8}$Al$_{6}$Ga$_{10}$Si$_{30}$&Sr(6d$\perp$)    &80 &ADP\cite{Shimizu}&\quad&Ba$_{8}$Cd$_{7.6}$Ge$_{38.4}$&Ba(6d$\parallel$)&58&ADP\cite{ChristensenDT2010}\\ \hline
Ba$_{8}$Ga$_{16}$Sn$_{30}$&Ba(6d$\perp$)    &67 &ADP\cite{SChristensen}&\quad&Ba$_{8}$Cd$_{7.6}$Ge$_{38.4}$  &Ba(6d$\perp$)    &78 &ADP\cite{ChristensenDT2010}\\ \hline
Ba$_{8}$Ga$_{16}$Sn$_{30}$&Ba(2a)           &81 &ADP\cite{SChristensen}&\quad&Ba$_{8}$In$_{16}$Ge$_{30}$  &Ba(6d$\parallel$)&65&ADP\cite{BentienPRB2005}\\ \hline
K$_{8}$Ga$_{8}$Sn$_{38}$&K(6d$\parallel$)&65 &ADP\cite{Tanaka}&\quad&Ba$_{8}$Ir$_{0.2}$Ge$_{43.2}$  &Ba(6d$\parallel$)&64&ADP\cite{FalmbiglInter2013}\\ \hline
K$_{8}$Ga$_{8}$Sn$_{38}$&K(6d$\perp$)    &97 &ADP\cite{Tanaka}&\quad&Ba$_{8}$Ir$_{0.2}$Ge$_{43.2}$  &Ba(6d$\perp$)    &85&ADP\cite{FalmbiglInter2013}\\ \hline
K$_{8}$Ga$_{8}$Sn$_{38}$&K(2a)           &112&ADP\cite{Tanaka}&\quad&Ba$_{8}$Pt$_{2.7}$Ge$_{41.7}$  &Ba(6d$\parallel$)&82&ADP\cite{MelnychenkoPRB2007}\\ \hline
K$_{8}$Zn$_{4}$Sn$_{42}$&K(6d$\parallel$)&78 &ADP\cite{XuJEM2017}&\quad&Ba$_{8}$Pt$_{2.7}$Ge$_{41.7}$&Ba(6d$\perp$)   &96&ADP\cite{MelnychenkoPRB2007}\\ \hline
K$_{8}$Zn$_{4}$Sn$_{42}$&K(6d$\perp$)    &97 &ADP\cite{XuJEM2017}&\quad&Ba$_{8}$Au$_{6}$Ge$_{40}$  &Ba(6d$\parallel$)&62 &ADP\cite{Johnsen}\\ \hline
K$_{8}$Zn$_{4}$Sn$_{42}$&K(2a)           &120&ADP\cite{XuJEM2017}&\quad&Ba$_{8}$Au$_{6}$Ge$_{40}$  &Ba(6d$\perp$)    &84 &ADP\cite{Johnsen}\\ \hline
Rb$_{8}$Sn$_{44}$&Rb(6d$\parallel$)&54 &ADP\cite{KaltzoglouJMC2008}&\quad&Ba$_{8}$Au$_{6}$Ge$_{40}$&Ba(2a)           &110&ADP\cite{Johnsen}\\ \hline
Rb$_{8}$Sn$_{44}$&Rb(6d$\perp$)    &81 &ADP\cite{KaltzoglouJMC2008}&\quad&Sr$_{8}$Ga$_{16}$Ge$_{30}$  &Sr(6d$\parallel$)&80 &ADP\cite{BentienPRB2005}\\ \hline
Rb$_{8}$Sn$_{44}$&Rb(2a)           &92 &ADP\cite{KaltzoglouJMC2008}&\quad&Sr$_{8}$Ga$_{16}$Ge$_{30}$  &Sr(6d$\perp$)    &104&ADP\cite{BentienPRB2005}\\ \hline
 \textbf{Skutterudite}& & & &\quad& & & & \\ \hline
NaFe$_{4}$Sb$_{12}$&Na &162 &ADP\cite{SchnellePRB2008}&\quad&CaFe$_{4}$Sb$_{12}$&Ca &114 &ADP\cite{SchnellePRB2008}\\ \hline
SrFe$_{4}$Sb$_{12}$&Sr &113 &ADP\cite{SchnellePRB2008}&\quad&BaFe$_{4}$Sb$_{12}$&Ba &126 &ADP\cite{SchnellePRB2008}\\ \hline
 \textbf{Pyrochlore}& & & &\quad& & & & \\ \hline
KOs$_{2}$O$_{6}$&K &61 &HC\cite{HiroiJPSJ2012}&\quad&RbOs$_{2}$O$_{6}$&Rb &66 &HC\cite{HiroiJPSJ2012}\\ \hline
CsOs$_{2}$O$_{6}$&Cs &75 &HC\cite{HiroiJPSJ2012}&\quad& &  & &\\
   \hline
   \hline
  \end{tabular}
  \label{SourceData}
  \end{center}
\end{table*}

\subsection{Van der Waals potential model and harmonic approximation}
%%van der Waals potential model and Rfree
To interpret the guest vibration modes, we introduce a modified Morse potential as we did previously.~\cite{Morse,WuPRB2016}
As shown in Fig.\,\ref{Fig1}, the potential of a guest atom inside a cage can be simplified by pair-wise potentials and expressed by,
\begin{equation}
\begin{split}
V_{t}(r)\,=\,&V(R+r)+V(R-r)\\
\,=\,&ae^{-nb(R+r-r_{e})}-ane^{-b(R+r-r_{e})}\\
&+ae^{-nb(R-r-r_{e})}-ane^{-b(R-r-r_{e})}
\end{split}
\label{potential}
\end{equation}
where $R$ is shown in Fig.\,\ref{Fig1} and can be viewed as a cage radius, $r_{e}$ is the equilibrium distance and can be estimated by $R_{\mathrm{guest}}$+$R_{\mathrm{host}}$ using van der radii of atoms; $n$, $a$ and $b$ are free parameters with $n\gg1$.
The expression can be simplified by introducing $R_{\mathrm{free}}$=$R$-$r_{e}$:
\begin{equation}
\begin{split}
V_{t}(r)\,=\,&ae^{-nb(R_{\mathrm{free}}+r)}-ane^{-b(R_{\mathrm{free}}+r)}\\
&+ae^{-nb(R_{\mathrm{free}}-r)}-ane^{-b(R_{\mathrm{free}}-r)}
\end{split}
\label{potential-simp}
\end{equation}
What we are focusing on is the potential minimum, where a guest atom is on the center and $r$ equals to zero.
One can easily derive the following equations,
\begin{equation}
%%\begin{split}
V_{t}(r=0)\,=\,2a(e^{-nbR_{\mathrm{free}}}-ne^{-bR_{\mathrm{free}}})
%%\end{split}
\label{zero}
\end{equation}
%%%=====================
\begin{equation}
%%%\begin{split}
V_{t}^{'}(r=0)\,=\,0
%%%\end{split}
\label{1st}
\end{equation}
%%%======================
\begin{equation}
%%%\begin{split}
V_{t}^{''}(r=0)\,=\,2anb^{2}(ne^{-nbR_{\mathrm{free}}}-e^{-bR_{\mathrm{free}}})
%%%\end{split}
\label{2nd}
\end{equation}
A potential minimum requires $V_{t}^{''}\,>\,0$, so the cage radius should satisfy the following criteria,
\begin{equation}
%%%\begin{split}
R\,<\,R_{c}\,=\,r_{e}+\frac{\ln(n)}{nb-b}.
%%%\end{split}
\label{rcondition}
\end{equation}
otherwise, the central position is not stable, and two potential minima arise near the center.

%%Harmonic approximation, Fc     %%relationship of Fc and Rfree
Under harmonic approximation, the force constant ($Fc$) of a harmonic oscillator can be expressed by $V_{t}^{''}(r=0)$, with the assumption that the repulsive term is much larger than the attractive term.
\begin{equation}
\begin{split}
\mathrm{Fc}&\,=\,2anb^{2}(ne^{-nbR_{\mathrm{free}}}-e^{-bR_{\mathrm{free}}})\\
&\,\cong\,2an^{2}b^{2}e^{-nbR_{\mathrm{free}}}
\end{split}
\label{Fcform2}
\end{equation}
Experimentally, $Fc$ can be estimated by the harmonic oscillator model that $Fc\,=\,m\omega_{E}^{2}$.
Therefore, the simple exponential relationship can be tested by experiment data.

\subsection{Unified picture of different ALE modes in cage compounds}
%%For type-I clathrate and parameters
As shown in Fig.\,\ref{Fig2}\,(b-c), the exponential behavior, $Fc\,=\,Ae^{-BR_{\mathrm{free}}}$, holds for each guest mode in type-I clathrate.
Here $A\,=\,2an^{2}b^{2}$ and $B\,=\,nb$, corresponding to the parameters derived earlier.
$R_{\mathrm{free}}$ was estimated using $R_{\mathrm{free}}$\,=\,$R$-$R_{\mathrm{guest}}$-$R_{\mathrm{host}}$, and $R$ was defined as the distance between a guest atom and its nearest neighbour.~\cite{WuPRB2014,WuPRB2016}
van der Waals radii were used for $R_{\mathrm{guest}}$ and $R_{\mathrm{host}}$~\cite{WuPRB2016,Mantina,Bondi} and they are: 1.40\,{\AA}(Sr), 1.57\,{\AA}(Ba), 1.01\,{\AA}(Na), 1.32\,{\AA}(K), 1.44\,{\AA}(Rb), 1.61\,{\AA}(Cs), 1.27\,{\AA}(Ca), 2.10\,{\AA}(Si), 2.11\,{\AA}(Ge), 2.17\,{\AA}(Sn), 1.87\,{\AA}(Ga), 1.39\,{\AA}(Zn), 1.4\,{\AA}(Cu), 1.63\,{\AA}(Ni), 1.72\,{\AA}(Ag), 1.66\,{\AA}(Au), 1.58\,{\AA}(Cd), 1.93\,{\AA}(In), 1.84\,{\AA}(Al), 1.52\,{\AA}(O) and 2.06\,{\AA}(Sb).
Irrespective of different vibration modes, the component elements are same for different cages, therefore we assume $nb$ keeps the same value for each case, and $a$ is variant depending on the effective number of guest-host pairs (the total potential is the sum of the component pairwise potentials).
In the fitting, we set $B\,=\,3.0$, which we reported previously~\cite{WuPRB2016}.
The fitting parameters $A$ and $B$ are listed in Table\,\ref{FitParameter}.
For the 6d$\parallel$ mode, the result from ADP data is consistent with the result from heat capacity data~\cite{WuPRB2016}.
The 2a and 6d$\perp$ modes show similar behavior but are much stronger than the 6d$\parallel$ mode.
The difference will be discussed later on, but it is noteworthy that it does not originate from space parameters, which have been already renormalized by $R_{\mathrm{free}}$.

\begin{figure*}[htbp]
\centering
\includegraphics*[width=0.9\linewidth]{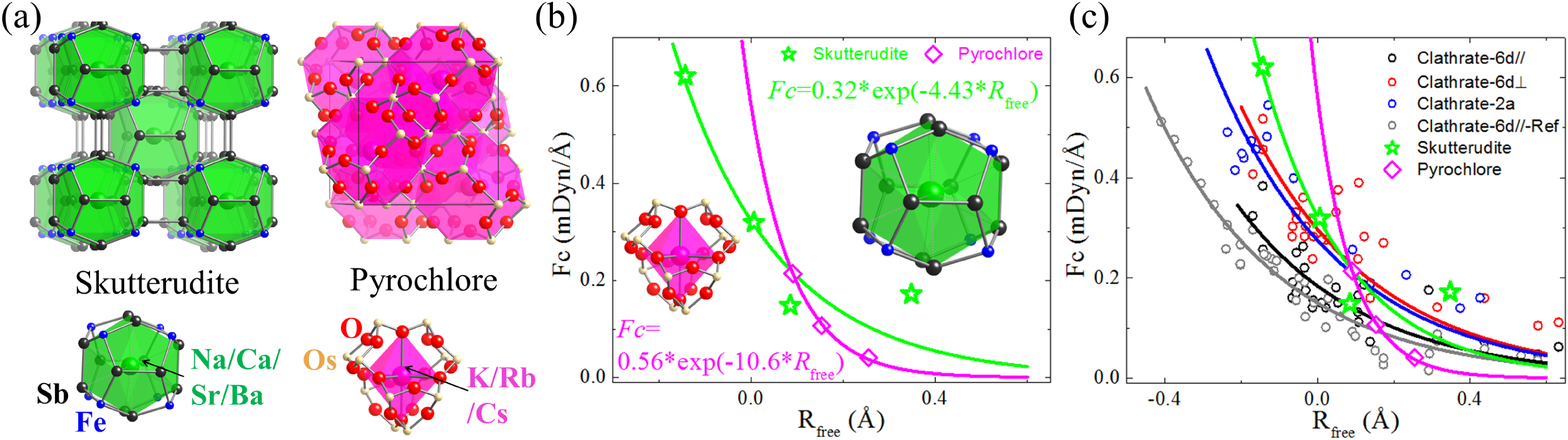}
\caption
{
(a)\,lattice structures of a typical skutterudite and pyrochlore, their cages are shown by green and pink colors, respectively.
The closest interspaces between guest and cage atoms are shown for both compounds.
(b)\,The relationship between Fc and $R_{\mathrm{free}}$ for different vibration modes of guest atoms in skutterudites and pyrochlores.
The lines are fitting results by employing exponential functions, and their colors are corresponding to the colors of cages.
The data of skutterudite and pyrochlore are derived from Refs.\,\cite{SchnellePRB2008,HiroiJPSJ2012}
(c)\,A summary for clathrates, skutterudites and pyrochlores.
}
\label{Fig3}
\end{figure*}

%%For skutterudite and pyrochlore and parameters
For skutterudite and pyrochlore, although lack of sufficient data point, the exponential dependence can also be clearly seen in Fig.\,\ref{Fig3}\,(b-c).
Obviously, the guest-host interaction is much stronger than that in type-I clathrate, this might be the reason that rattling vibrations in some of these compounds have much higher energies and can not be clearly distinguished from other high energy non-dispersive phonon modes.~\cite{KozaPRB2010,MelotPRB2009}

%%%================================================================
\begin{table}[htbp]
\centering
\caption
{
Fitting parameters of the relationship between $Fc$ and $R_{\mathrm{free}}$.
$Fc\,=\,Ae^{-BR_{\mathrm{free}}}$.
}
\begin{center}
 \begin{tabular}{cccccccc}
 \hline \hline
     vibration modes&A&B                                                                   \\ \hline
     clathrate-2a                                     &0.28            &3.0                \\
     clathrate-6d$\perp$                              &0.30            &3.0                \\
     clathrate-6d$\parallel$                          &0.18            &3.0                \\
     clathrate-6d$\parallel$-ref\,\cite{WuPRB2016}    &0.15            &3.0                \\
     Skutterudite                                     &0.32            &4.4                \\
     Pyrochlore                                       &0.56            &10.6               \\  \hline \hline
  \end{tabular}
\end{center}
\label{FitParameter}
\end{table}
\section*{Discussion}
%%interaction resolution in type-I clathrate
To have a deep insight into the guest-host interactions based on the van der Waals potential model, we consider the nearest cage atoms as primary atoms that strongly interact with the guest atom.
The closest interspace for each cage is shown in Fig.\,\ref{Fig2} and Fig.\,\ref{Fig3}.
In type-I clathrate, for both dodecahedral and tetrakaidecahedral cages, guest atoms have eight nearest neighbors; however the closest interspaces are in different shapes: a cube and a twisted cuboid, respectively.
Therefore the 2a mode inside the cube is isotropic, while the 6d mode inside the twisted cuboid is anisotropic and splits into a 6d$\parallel$ mode and a 6d$\perp$ mode.
The total guest-host interactions inside the tetrakaidecahedral cage can be resolved along 6d$\parallel$ and 6d$\perp$ directions, as shown by the arrows in Fig.\,\ref{Fig2}\,(b-c).
According to the geometry of the twisted cuboid, the strength of the interaction along 6d$\parallel$ direction is weaker than that along 6d$\perp$ direction, and the ratio is around 0.62, which is very close to the ratio of the fitting parameters (shown in table\,\ref{FitParameter}), A$_{\mathrm{6d}\parallel}$/A$_{\mathrm{6d}\perp}$\,$\cong$\,0.6.
This is not just a coincidence, but rather indicates that the proposed van der Waals interaction works well for the rattling system, and the interaction is closely associated with the effective number of pairwise potentials (or the coordination number) and the cage geometry.

%%chemical environment for skutterudite and pyrochlore.---> The formation of clathrate, pyrochlore and skutterudite.
%%comment on the off-center behavior
In skutterudite and pyrochlore, the situation is very different, not only in the number and geometry of the primary cage atoms, but also in the component elements.
Correspondingly, the fitting parameters $A$ and $B$ are very different.
The cage of clathrate is mainly composed of Si/Ge/Sn, while the cage of skutterudite is mainly composed of P/As/Sb and the cage of pyrochlore is mainly composed of O.
The elements in the later cases have much large electronegativity, which may introduce additional interactions, such as ionic interactions, in addition to the principal van der Waals interactions.
Under such an assumption, the guest atoms should be more easily to be off-centered, and this is true for pyrochlore compounds,~\cite{ShoemakerPRB2011,HiroiJPSJ2012} while skutterudite cages are usually very small, so off-centered guest atoms were rarely observed.
This conclusion can be supported by the critical radius estimated earlier as shown in equation \ref{rcondition}.
The large $nb$(B) values for skutterudite and pyrochlore would give rise to a small $R_{c}$, and it is readily for a cage to be oversized and the guest atom becomes off-centered.

\section*{Conclusions}
%%%%%%%%%%%%%%%%%%%%%%%%%%%%%%%%%%%%%%%%%%%%
In the present work, we studied the ALE vibration modes of guest atoms in cage materials by focusing on type-I clathrates, skutterudites and pyrochlores.
We showed that the 3D van der Waals-type interaction, which is usually important only in molecular solids but negligible in the other types of solids (ionic, covalent, etc.), can be clearly observed in cage compounds with strong covalent framework.
%%Medlung potential sum to be zero
The strong covalent bonded cages create a solid wall and chemical pressure between guest atoms and their surrounding frameworks, stabilizing the anomalous low energy phonon modes.
In addition to free space and mass parameters, which we derived previously~\cite{WuPRB2016}, we introduced another three chemical environment associated factors:
(1) coordination number of guest atoms, or the effective number of pairwise potentials;
(2) geometry of cages;
(3) electronegativity of the component elements of a cage.
These five factors work together to determine the interesting soft modes of the rattling vibrations.
The 3D van der Waals-type interaction in cage materials should be highly evaluated in the community of superconductivity and thermoelectricity.

\section*{Acknowledgment}
This work was partially supported by the AIMR collaborative research program.
JX acknowledges the financial support by the National Nature Science Foundation of China (NSFC No. 11304327).

\label{Bibliography}
\bibliography{ALEMode} % The references (bibliography) information are stored in the file named "Bibliography.bib"
\bibliographystyle{unsrtnat} % Use the "unsrtnat" BibTeX style for formatting the Bibliography

\end{document}